\documentclass[conference,a4paper,twocolumn]{IEEEtran}

\IEEEoverridecommandlockouts

\ifCLASSINFOpdf
\else
\fi
  \usepackage{tipa}
  \usepackage[pdftex]{graphicx}
	\usepackage{epsfig}
	\usepackage{epstopdf}
\usepackage{subcaption}

\usepackage[cmex10]{amsmath}
\begin{document}



\title{Deep Receiver Design for Multi-carrier Waveforms Using CNNs}


\author{\IEEEauthorblockN{
Yasin YILDIRIM}
\IEEEauthorblockA{
Istanbul Technical University \\
Istanbul, Turkey \\
yildirimy17@itu.edu.tr
}
\and

\IEEEauthorblockN{
Sedat \"OZER}
\IEEEauthorblockA{
Bilkent University\\
Ankara, Turkey\\
sedat@cs.bilkent.edu.tr}

\and

\IEEEauthorblockN{
Hakan Ali \c{C}IRPAN}
\IEEEauthorblockA{
Istanbul Technical University\\
Istanbul, Turkey\\
cirpanh@itu.edu.tr}

\thanks{This work was partially supported by TUBITAK's 2232: International Fellowship for Outstanding Researchers program (Project No: 118C356), however, that support does not necessarily mean that TUBITAK also supports all the content and comments made in this paper.}

}


%


\maketitle




\begin{abstract}
In this paper, a deep learning based receiver is proposed for a collection of multi-carrier wave-forms including both current and next-generation wireless communication systems. In particular, we propose to use a convolutional neural network (CNN) for jointly detection and demodulation of the received signal at the receiver in wireless environments. We compare our proposed architecture to the classical methods and demonstrate that our proposed CNN-based architecture can perform better on different multi-carrier forms including OFDM and GFDM in various simulations. Furthermore, we compare the total number of required parameters for each network for memory requirements.
\end{abstract}


%
\IEEEpeerreviewmaketitle


\section{Introduction}

Modern  wireless communication systems heavily rely on multi-carrier wave-forms. Orthogonal Frequency Division Multiplexing (OFDM) \cite{chang1966synthesis}, Generalized Frequency Division Multiplexing (GFDM) \cite{fettweis2009gfdm}, Filter Bank Multi-Carrier (FBMC) and Universal Filtered Multi-Carrier (UFMC) \cite{schaich2014waveform} are examples for multi-carrier wave-forms. Such variability in wave-forms, brings up the need for using unified receiver architectures that are flexible enough to be used for different multiple wave-forms. For example, a receiver that can be used for both OFDM and GFDM systems would be a great use for wireless communication systems. Deep learning (DL) based receiver solutions are such emerging potential solutions in receiver design and there are already deep learning based solutions proposed in the literature as in \cite{turhan2019deep,van2019deep}.

Among many deep learning architectures, Convolutional Neural Networks (CNNs) have revolutionized multiple research fields (as in computer vision and natural language processing). That is due to the fact that CNNs provide significant performance improvements, when compared to the classical approaches in many applications. However, their use in the receiver design for wireless communication remains limited. The literature for the receiver design mainly focused on designing deep architectures using fully-connected (dense) layers, however, there are not many works using 2D CNNs in the relevant literature. CNNs showed their value on spatial (2D) datasets already. In communication systems, it is essential to deal with complex numbers and complex numbers can also be represented and considered as 2D (spatial) data. Consequently, a 2D CNN dealing with spatial relations among the data points can be useful in the receiver design. 

In this paper, we mainly analyze and report the performance of 2D CNNs in receiver design for wireless systems using multi-carrier wave-forms. Among multiple carrier types, we focus on studying the performance of 2D CNNs on OFDM and GFDM systems.

Classical GFDM receivers are known with their heavy computational requirements \cite{kislalsurvey}. While deep learning architectures can be considered as alternatives for receiver design, as mentioned above, they also introduce their own complexities. The existing architectures using fully connected layers typically require heavier computational complexities, when compared to the use of typical 2D CNNs. Using 2D convolutional layers typically  provides a significant reduction in computational complexity when they replace the larger sized fully connected layers containing the largest number of neurons in the earlier layers; thus they can help reduce the complexities of the used deep architectures.

 \addtolength{\topmargin}{+.075in}
 
In this paper, our contributions are two-fold: (i) To our best knowledge, this is the first work that introduces data detector without channel equalization using only a deep 2D CNN architecture by eliminating the need for using a coarse detector for multicarrier systems; and (ii) we introduce using CNNs to reduce the required complexities in deep architectures and provide an analysis on parameter computation for various deep architectures.  

 \begin{figure*}[!ht]
	\centering
	\vspace{0.1in}
	\includegraphics[width=0.7\linewidth]{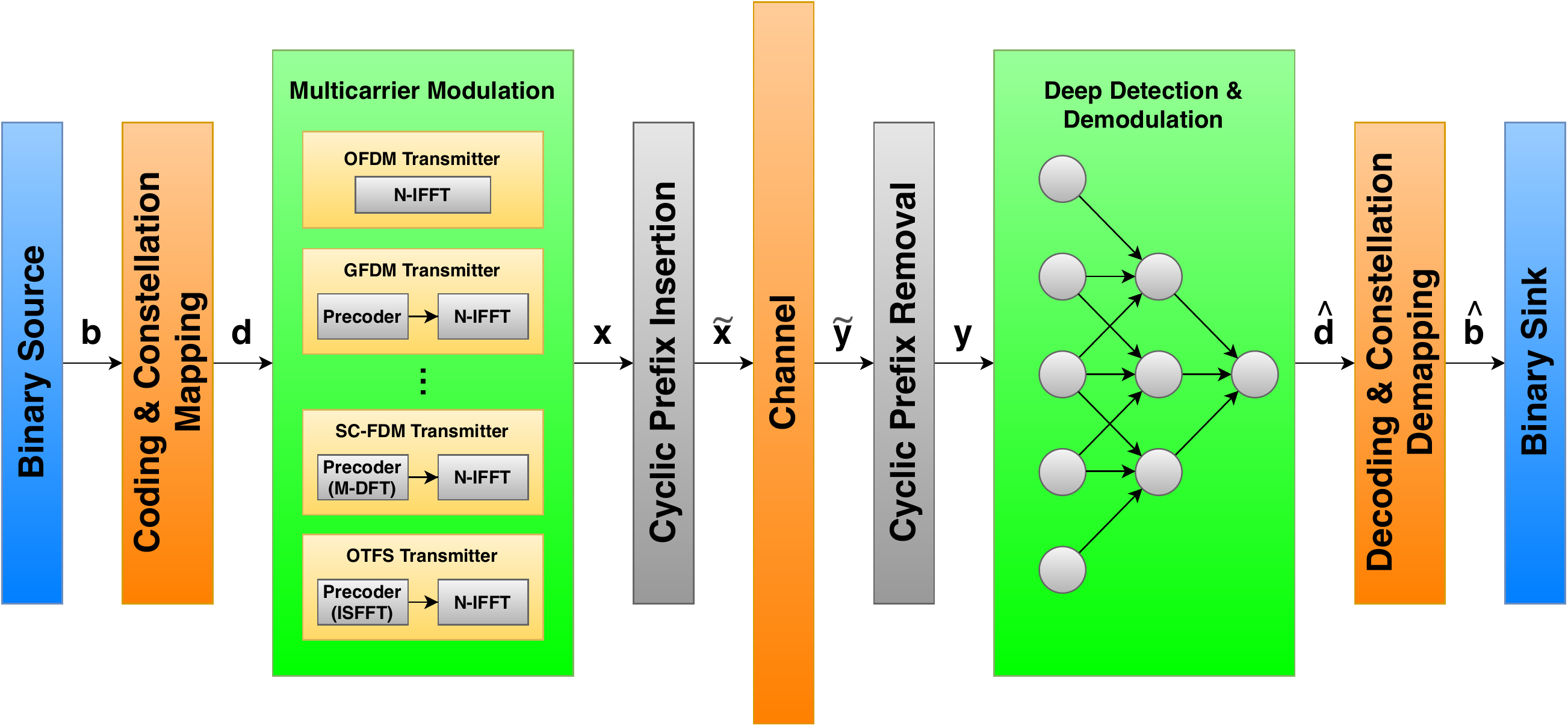}	
	\caption{An overview of the proposed deep receiver architecture.}
	\label{fig:mc}
\end{figure*}

\section{Related Work}

Deep learning-based architectures have been studied in receiver design in the recent literature, however, their properties yet to be fully exploited. While the researchers started to focus on including deep network architectures in wireless communication systems, the main focus has been on utilizing fully-connected (dense) networks in various communication systems. However, using convolutional networks  (2D or 3D) can reduce the computational cost when compared to dense networks, while providing similar or better performance to that of dense networks as shown in this paper (see Table \ref{tab:trainpar}). Dense networks are also known as multi-layer perceptron (MLP) in some literature and in this paper, we will use both terms interchangeably to refer to the same network type. 

Fully connected (dense) networks were used in \cite{dorner2017deep} as a part of a communication system. In \cite{ye2017power}, the authors proposed using dense networks to design receivers for OFDM-based systems. In \cite{he2018model,zhang2019artificial} the authors proposed an OAMP based algorithm that uses training to learn the OAMP parameters in MIMO systems. In \cite{van2019deep}, a dense network was proposed as detector. In \cite{balevi2019one}, the authors proposed using a dense network for OFDM receivers under the constraint of one-bit complex quantization. Long-short term memory (LSTM) based networks were also used in communication systems. For example, in \cite{gao2018comnet}, a deep architecture using LSTM and fully connected layers for OFDM systems was proposed. In \cite{fang2017deep}, deep belief networks and auto-encoders were proposed. Auto-encoders were also proposed in \cite{kim2018deep}. Similarly, in \cite{o2017deep}, a fully-connected auto-encoder structure was also proposed. 

 \addtolength{\topmargin}{+.075in}

Most of the above-mentioned work focused on utilizing dense networks in different architectures. In \cite{farsad2017detection}, the authors studied the performance of dense networks, convolutional networks and recurrent neural networks for chemical (molecular) communication systems. In \cite{zhao2018deep}, the authors proposed an auto-encoder architecture utilizing 2D convolutions for OFDM.
 
 The closest work to ours in the literature is the work in \cite{turhandeep,turhan2019deep} as they both design a deep receiver for GFDM system using 2D CNNs. The receiver design in \cite{turhandeep,turhan2019deep} contains  two  detectors: a coarse detector and a fine detector where the coarse  detector uses one of the classical methods (e.g., zero force or Minimum Mean Square Estimator) first and then a 2D CNN is used to further improve the detection performance. 
 
 In this work, we investigate on utilizing only a deep architecture that combines both coarse and fine detectors for multi-carrier wave-forms. By representing a complex number as 2D number, we can utilize 2D CNNs in the receiver. Therefore, in this paper, we study the performance of 1D and 2D CNNs and compare that to dense networks for communication systems utilizing multi-carrier wave-forms. Furthermore, while the literature using  deep architectures mostly focused on OFDM wave-forms, in this work, we study the performance of using various deep architectures for multiple systems including OFDM and GFDM.

 \section{System Architecture: Overview}
 
In this section, we provide an overview of a wireless system using multi-carrier wave-forms. Fig.~\ref{fig:mc} shows the block diagram of such wireless system \cite{anwarperformance}.  

Binary data vector $\textbf{b}$ is generated by a data source. Coding and $2^\phi$-valued complex constellation mapping is performed to obtain symbol vector $\textbf{d}$, where $\phi$ is the modulation order. The resulting vector $\textbf{d}$ has block-based structure. Thus, it can be decomposed in frequency and time space into $M$ subsymbols and $K$ subcarriers, wherein $K$ is the total number of subcarriers and $M$ is the number of symbols in one block according to $\textbf{d}=(\textbf{d}_0,\dots,\textbf{d}_{M-1}^{\; T})^T$ and $\textbf{d}_m=(\textbf{d}_{0,m},\dots,\textbf{d}_{K-1,m})^T$. The total number of symbols  ($N$) in one multi-carrier wave-form symbol becomes $N = KM$. $N$ dimensional $\textbf{x}$ time vector is created by modulating matrix $\textbf{d}$ with a desired multi-carrier wave-form (e.g., OFDM, GFDM, SC-FDM, SC-FDE or OTFS) where $x(n)$ is defined as:

 \begin{equation}
x(n)=\sum_{k=0}^{K-1}\sum_{m=0}^{M-1}d_{k,m}g_{k,m}(n),\quad n = 0,\dots,N-1.
\end{equation}
 
The model of $g_{k,m}(n)$ vector differs according to the preferred wave-form type. For example, in GFDM, the $g_{k,m}(n)$ is defined as in Eq.~\ref{eq:gkmn} where, operations such as pulse shaping, upsampling and frequency shifting are applied on vector \textbf{d}.

\begin{equation}
g_{k,m}(n)=g\big((n-mK){modN}\big)exp\Big(j2\pi\frac{k}{K}n\Big),
\label{eq:gkmn}
\end{equation}
where $g_{k,m}(n)$ is formed with the  pulse shaping filter $g(n)$. The linear relationship between the vectors $\textbf{x}$ and $\textbf{d}$ is given as follows: $\textbf{x}=\textbf{A}\textbf{d}$. The $NxN$ modulation matrix $\textbf{A}$ is created as shown in Eq. \ref{eq:modmat} with $g_{k,m}(n)$ which defines the columns of $\textbf{A}$. The mapped bits in $\textbf{d}$ are modulated by the modulation matrix $\textbf{A}$. Here, the modulation matrix is selected according to the chosen modulation type:

\begin{equation}
\textbf{A}=[g_{0,0},\ldots,g_{K-1,0},g_{0,1},\ldots,g_{K-1,1},\ldots,g_{K-1,M-1}]
\label{eq:modmat}
\end{equation}

Cyclic prefix is added to the resulting modulated signal to cope with multipath channel effects and adding cyclic prefix yields the time vector $\Tilde{\textbf{x}}$ to be transmitted. The received signal is exposed to a frequency selective Rayleigh fading channel as considered in Eq.~\ref{eq:received}:
 
 \begin{equation}
 \Tilde{\textbf{y}}=\Tilde{\textbf{H}}\Tilde{\textbf{x}} + \Tilde{\textbf{w}}
 \label{eq:received}
 \end{equation}
 where, we assume that the channel length is shorter than the CP and perfect synchronization is ensured. After removing the cyclic prefix,  Eq.~\ref{eq:received} can we re-written as in Eq.~\ref{eq:rec}:

 \begin{equation}
 \textbf{y}=\textbf{H}\textbf{x} + \textbf{w} 
 \label{eq:rec}
 \end{equation}
where, $\textbf{y}$ is $N$ dimensional and the received matrix $\textbf{H}$ is the $NxN$ circular convolution matrix constructed from the channel impulse response coefficients given by $\textbf{h} = [h(1),h(2),\dots,h(N_{ch})]^T $,
and $\textbf{w}$ is $N$ dimensional additive white Gaussian noise (AWGN) vector. The elements of $\textbf{h}$ and $\textbf{w}$ follow $\mathcal{C}\mathcal{N}(0,1)$ and $\mathcal{C}\mathcal{N}(0,\,\sigma_w^{2})$ distributions respectively, where $\mathcal{C}\mathcal{N}(\mu,\,\sigma^{2})$ shows the distribution of a circularly symmetric complex Gaussian random variable with mean $\mu$ and variance $\sigma^2$. Combining Eq.~\ref{eq:rec} with the linear relation of $\textbf{x}=\textbf{A}\textbf{d}$ yields to Eq.~\ref{eq:r1}:

 \begin{equation}
 \textbf{y}=\textbf{H}\textbf{A}\textbf{d} + \textbf{w} 
 \label{eq:r1}
 \end{equation}
 
 After zero forced (ZF) channel equalization we obtain the following equation:
 
 \begin{equation}
 \textbf{z}=\textbf{H}^{-1}\textbf{H}\textbf{A}\textbf{d} + \textbf{H}^{-1}\textbf{w} 
 \label{eq:r2}
 \end{equation}
 
 Linear demodulation of the signal can be
expressed as:
\begin{equation}
\hat{\textbf{d}}=\textbf{B}\textbf{z}
\label{eq:receiver}
\end{equation}
where $\textbf{B}$ is $NxN$ dimensional receiver matrix. Different types of linear detectors, e.g. zero forced receiver $\textbf{B}_{\textbf{ZF}}=\textbf{A}^{-1}$,matched filter (MF) receiver
$\textbf{B}_{\textbf{MF}}=\textbf{A}^{\textbf{H}}$ and minimum mean square error (MMSE) receiver $\textbf{B}_{\textbf{MMSE}}=(\textbf{R}_{\textbf{w}}+\textbf{A}^{\textbf{H}}\textbf{A})^{-1}\textbf{A}^{\textbf{H}}$
can be used to detect the data symbols from the equalized observation signal, where $\textbf{R}_{\textbf{w}}$ denotes the covariance matrix of the noise.

For neural network-based detection, the received signal $\textbf{y}$ is fed as an input to a deep neural network. After the estimated symbol vector $\hat{\textbf{d}}$ is taken from the output of the deep receiver, decoding and constellation demapping operations are performed. The resulting vector $\textbf{b}$ is sent to the binary sink.

\begin{figure}[t!]
	\centering
	\vspace{0.1in}
	\includegraphics[width=1.00\linewidth]{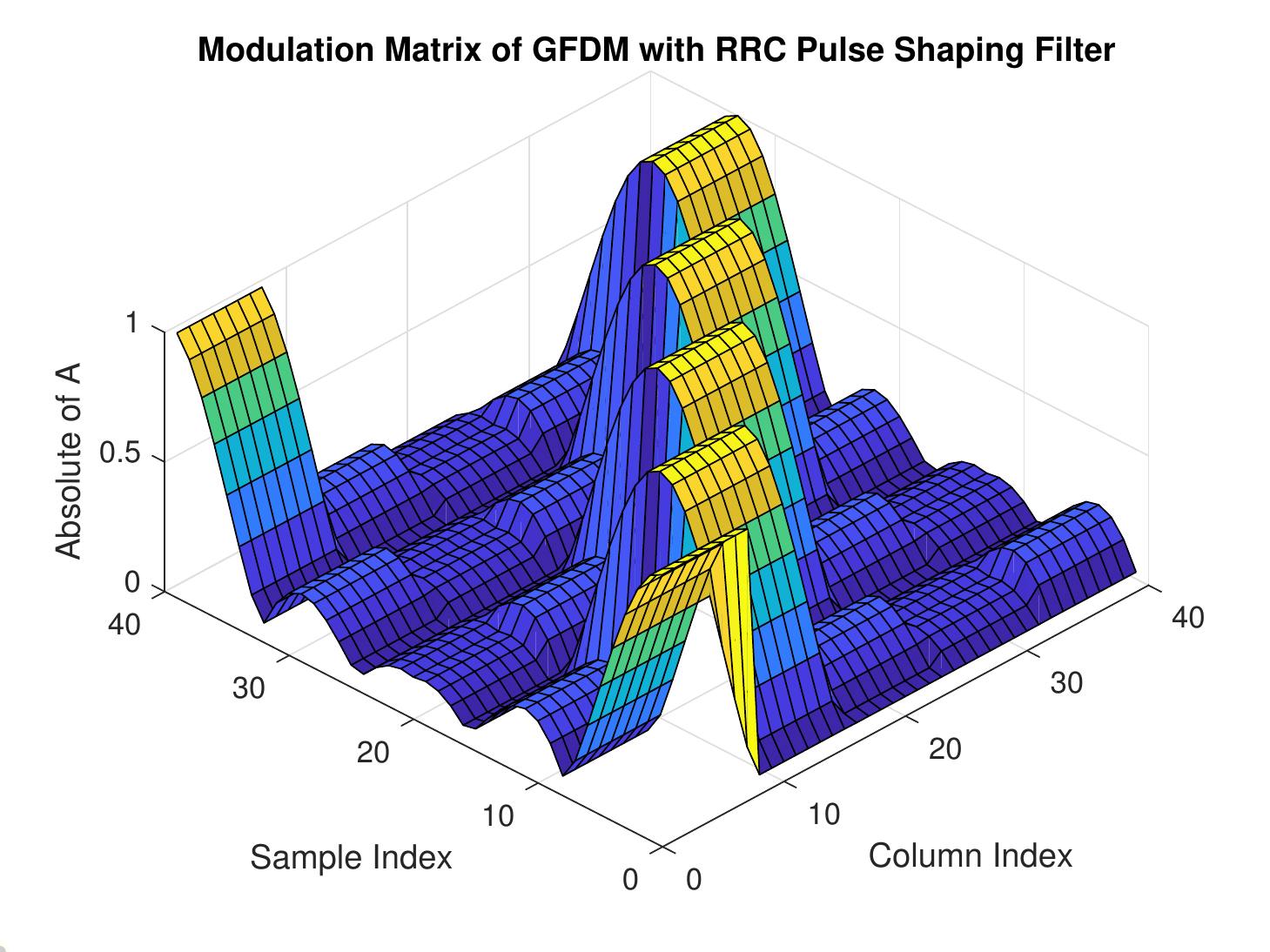}	
	\caption{Visualization of the GFDM Modulation Matrix ( K=8, M=5, RRC Filter (\emph{a}=0.1))}
	\label{fig:modmatrix}
\end{figure}

Modulation matrix performs different operations for each wave-forms. OFDM modulation matrix performs $N$-point IFFT operation which means frequency shifting. GFDM modulation matrix performs upsampling, pulse shaping and frequency shifting. Furthermore, it has circular structure and because of that, it allows the use of cyclic prefix to make frequency domain equalization possible. The magnitude of the GFDM modulation matrix is depicted in Fig. \ref{fig:modmatrix}.

Next section focuses on the details of utilizing a deep neural network in the receiver part.

 \section{Deep Receiver with CNNs}
 
In wireless communication, signal detection can be considered as a classification process of recovering the transmitted signal from the  (distorted) received signal where a deep architecture can be used for classification in the deep receiver.

Wireless channel can be evaluated unchanged at small intervals. If the channel is known in small intervals, the transmitted symbols can be recovered. In classical receiver architecture, pilot signals are often used and the channel is estimated by using those pilot signals. Once the channel is estimated, channel equalization is performed to eliminate multipath effects. The deep learning technique has the same framework. Deep receiver works in two different modes: Training and Testing. Training mode can be considered as using the pilot signals which can relate to the training data. Using the training data, the model weights are obtained in the neural network to recover the symbols. Testing does not require any pilot symbol and works for recovering messages to be transmitted by using weights obtained in the training. 

A typical CNN may contain convolutional layers and fully connected (FC) layers. Our proposed CNN architectures are given in Fig. \ref{fig:mod} for both OFDM and GFDM. In our proposed deep receiver, we use multiple convolutional layers followed by FC layers. Based on the number of used convolutional layers and the number of FC layers, the performance of the deep receiver changes (see Section \ref{Simulations}). Complex input data is represented as a 2D vector (where each of real and imaginary parts of a complex number forms one dimension). 2D convolution is applied to the input data and deep features are extracted. At the end of the fully connected layers, the symbols are estimated. The vector $\textbf{d}$ mapped and coded is used as ground truth while the network is being trained. We used Adam optimizer to train all the networks. Mean Squared Error (MSE) loss function is used in our network during the training as in Eq. \ref{eq:loss}:

\begin{equation}
\mathcal{L}=\frac{1}{N}(\hat{\textbf{d}}-\textbf{d})^{T}(\hat{\textbf{d}}-\textbf{d})
\label{eq:loss}
\end{equation}

At the end of the process, it is expected that underlying relationship formed by wireless channel is learned and tried to be eliminated from the signal.

\section{Simulations} \label{Simulations}   

In this section, we simulate the wireless system given in Fig. \ref{fig:mc} and study the performance of the deep receiver for multi-carrier wave-forms. In particular, we study the performance for both OFDM and GFDM wave-forms separately. We compared the deep receiver performance with the traditional methods in terms of bit-error rates (BERs) under different signal-to-noise ratios (SNRs). For comparison, we compared deep receiver's performance to classical methods including matched receiver  with channel equalization, zero forced receiver without channel equalization and MMSE receiver  with channel equalization  \cite{michailow2014generalized}.  

 \begin{figure}[t!]
	\centering
	\includegraphics[width=1.0\linewidth]{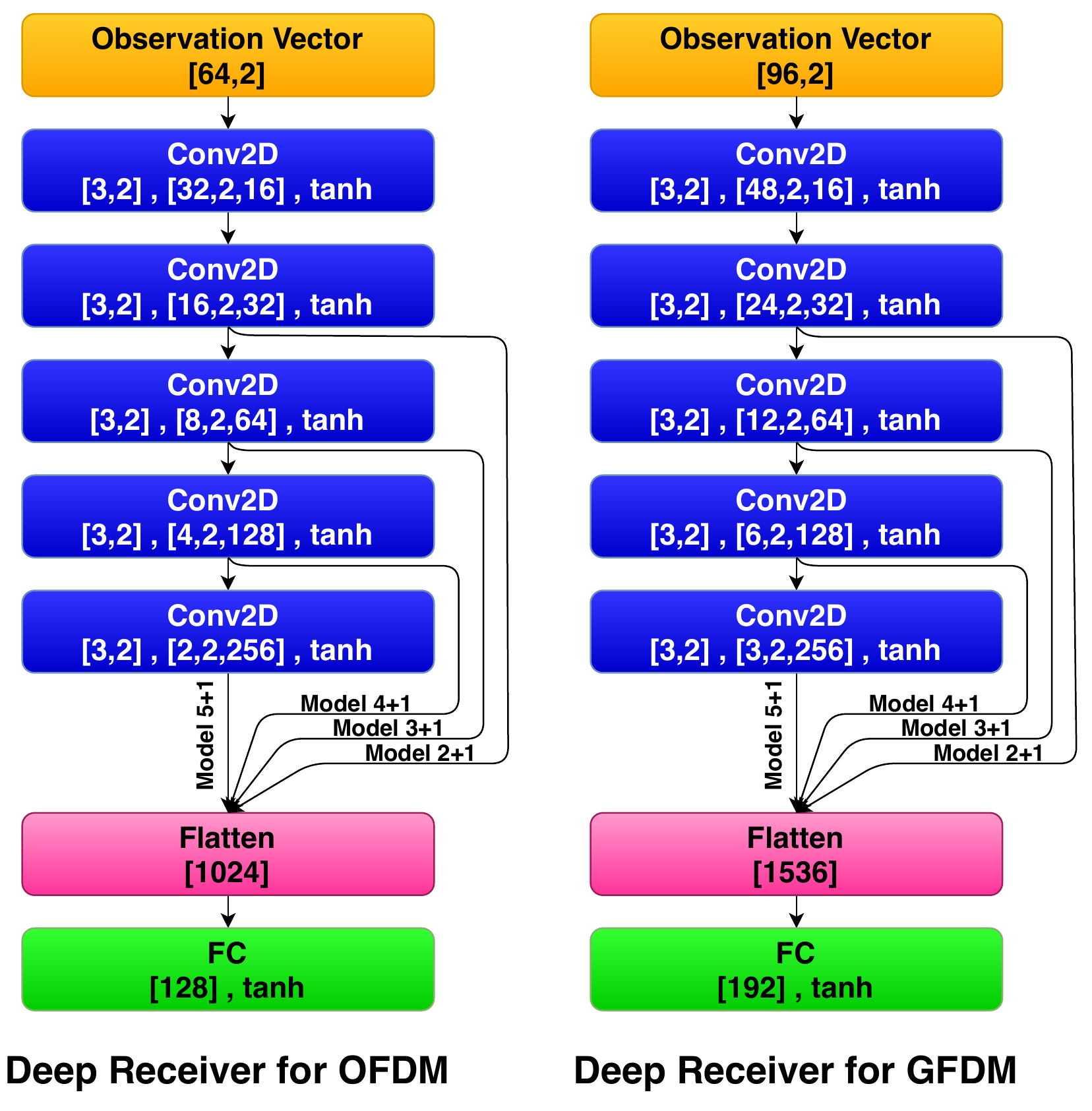}	
	\caption{Model Summaries for both OFDM and GFDM}
	\label{fig:mod}
\end{figure}

 \addtolength{\topmargin}{+.075in}

In our experiments, we use 10 tap Rayleigh fading with Extended Pedestrian A (EPA) channel model \cite{EPA} and that is fixed for both training and testing steps. In our all simulations, we used BPSK modulation and omitted channel coding and its corresponding decoding processes. In the following  subsections, we first present the experimental results for the OFDM wave-form and then present the results for the GFDM wave-form.

\begin{table}[t]
\caption{Simulation Parameters for Both OFDM and GFDM }
\begin{center}
\begin{tabular}{|c|c|c|c|}
\hline
\multicolumn{4}{|c|}{\textbf{OFDM \& GFDM Parameters used in the Simulations}} \\
\cline{1-4} 
\textbf{\textit{Description}}& \textbf{\textit{Parameter}}& \textbf{\textit{OFDM Value}}& \textbf{\textit{GFDM Value}} \\
\hline
Number of subcarriers&K & 64&32 \\
\hline
Number of subsymbols&M& 1&3 \\
\hline
Pulse shaping filter&g& -&RRC\\
\hline
Roll-off factor&$a$& -& 0.1 \\
\hline
Length of cyclic prefix&Ncp&16&24\\
\hline
Channel Taps (Fixed)&Nch&10&10\\
\hline
\end{tabular}
\label{gfdm}
\end{center}
\end{table}
\vspace{0.14in}

For both OFDM and GFDM experiments, we performed an empirical study by changing various hyperparameters of the deep architecture used in the receiver. Such as, number of filters (16, 32, 64, 128, 256, 512, 1024) for each layer, filter size (1-8), activation function (ReLU, Sigmoid and Tanh) of each layer except the output layer and the stride value (1, 2, 3). We choose the best configuration found in those experiments for the final parameters of the network and choose hyperbolic tangent (tanh) function as the activation function in the last layer as its output range is between -1 and 1. We also perform a search on the total number of needed convolutional layers by varying the total number of convolutional layers between 2 and 5. After the convolutional layers, we used a fully connected (FC) layer as the output layer. We named those models as 2+1, 3+1, 4+1 and 5+1, where 2+1 means that 2 convolutional layers and one FC layer as output is used. Similarly, 3+1 means that 3 convolutional layers and one FC layer is used in that network and so on. A summary of those networks is given in Fig. \ref{fig:mod} for CNNs where the given values in each convolution block represent the kernel size, the output size and the activation function respectively as used in that layer. All four models (2+1, 3+1, 4+1 and 5+1 models) are shown in the same figure. For instance, for the 3+1 model, the output after the first 3 convolution layers is connected to the flattening layer. Training step for each model is carried out separately at each SNR level and the networks trained up to 30 epochs with early stopping criteria.

Real and imaginary components of receiving signal are concatenated serially and given to the input of MLP and 1D CNN network. 5 different layers were chosen for the MLP. Layers have the neuron number 256, 128, 64, 32 and 16 respectively. The output layer consists of 192 neurons for GFDM 128 neurons for OFDM. Likewise, a structure consisting of 5 different layers is designed for 1D CNN. Convolution layers consist of 16, 32, 64, 128, 256 filters of size 3 respectively. The output layer consists of 192 neurons for GFDM 128 neurons for OFDM.

\begin{figure*}[t]

\begin{subfigure}{.33\linewidth}
  \centering
  \includegraphics[width=\linewidth]{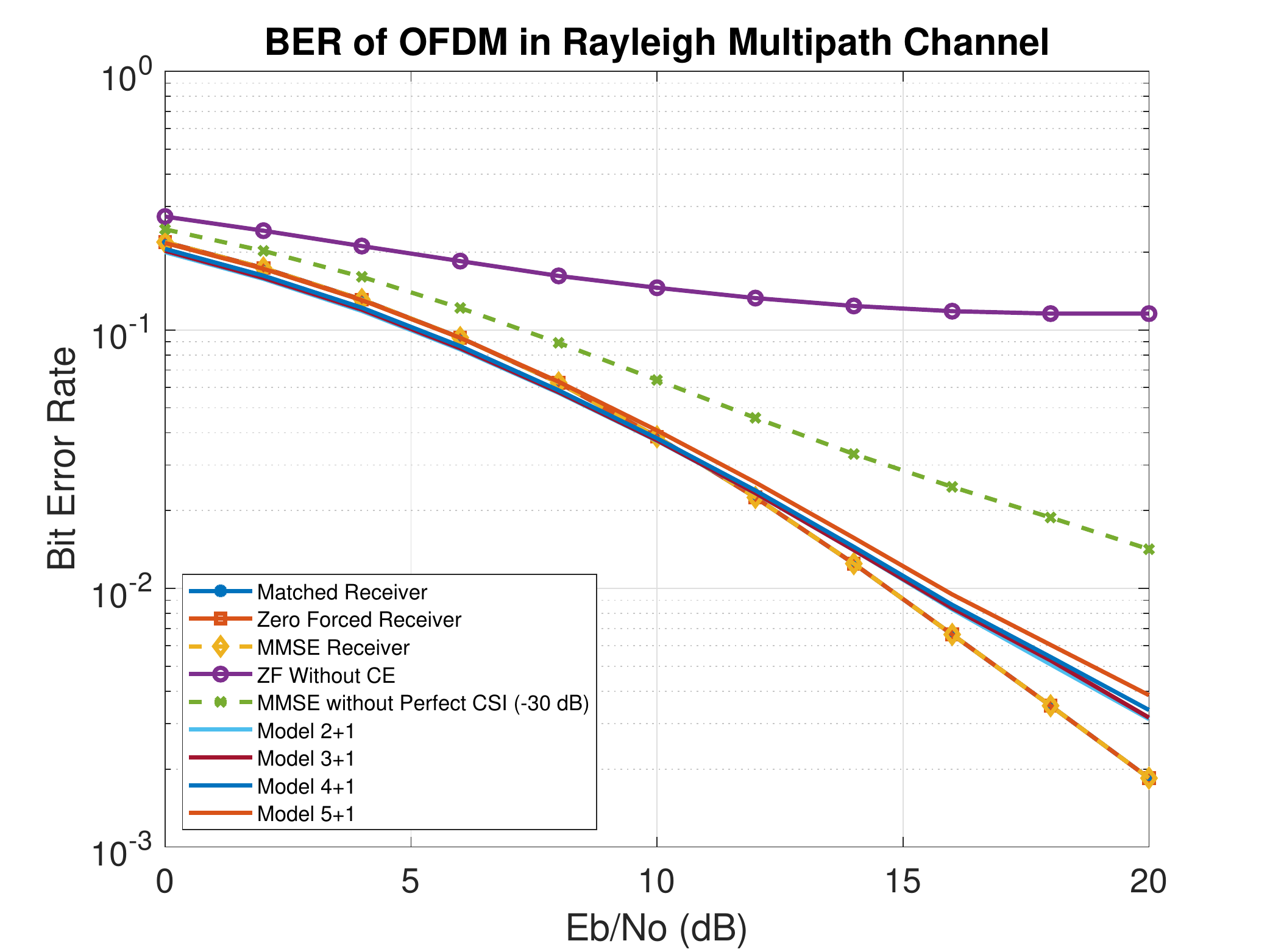}
    \vspace{-0.24in}
  \caption{\small OFDM results for 2D-CNNs}
  \label{fig:ofdm}
\end{subfigure}
~
\begin{subfigure}{.33\linewidth}
  \centering
  \includegraphics[width=\linewidth]{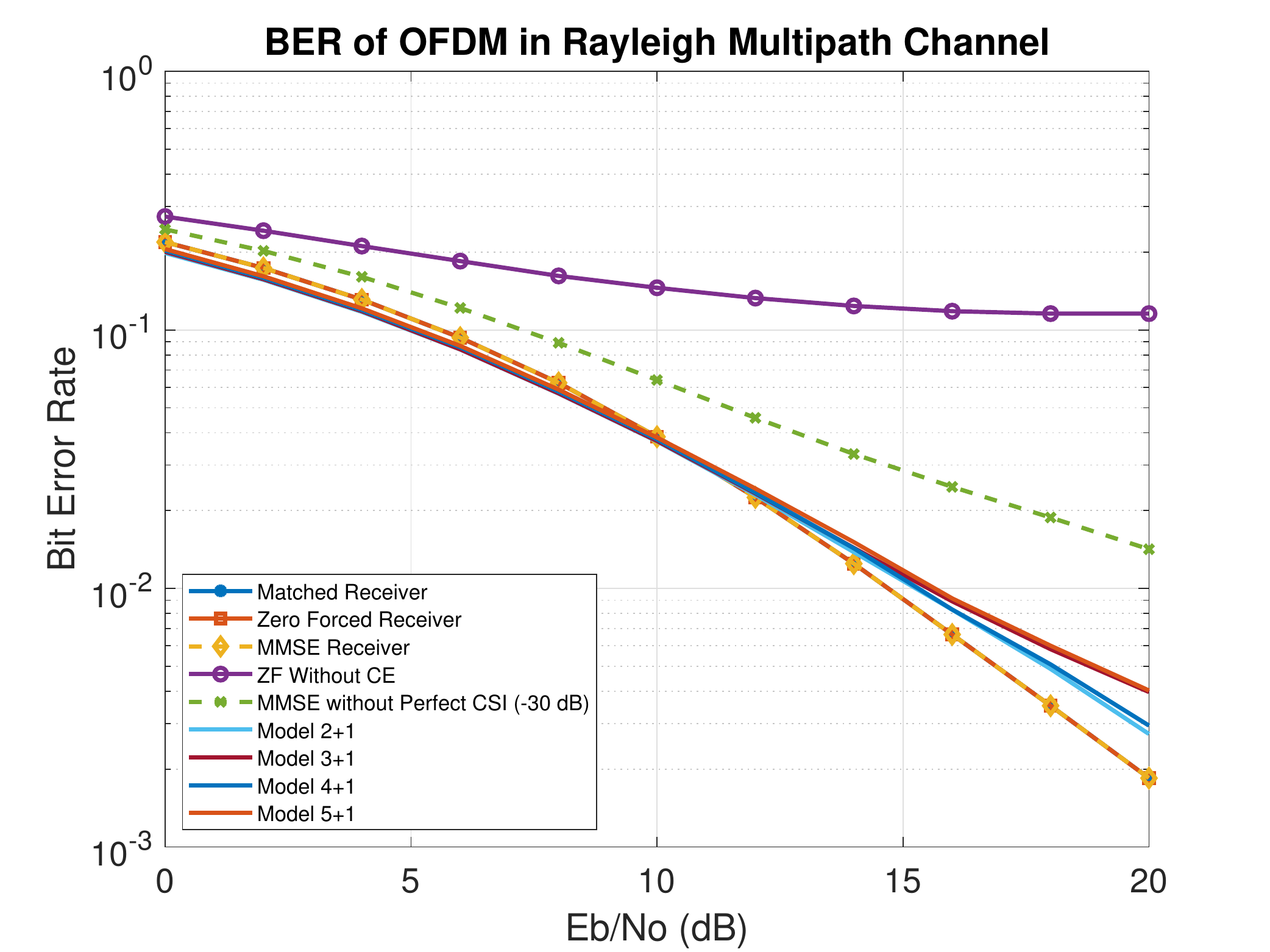}
    \vspace{-0.24in}
  \caption{\small OFDM results for 1D-CNNs}
  \label{fig:OFDM-1Dconv}
\end{subfigure}%
~
  \begin{subfigure}{.33\linewidth}
  \centering
  \includegraphics[width=\linewidth]{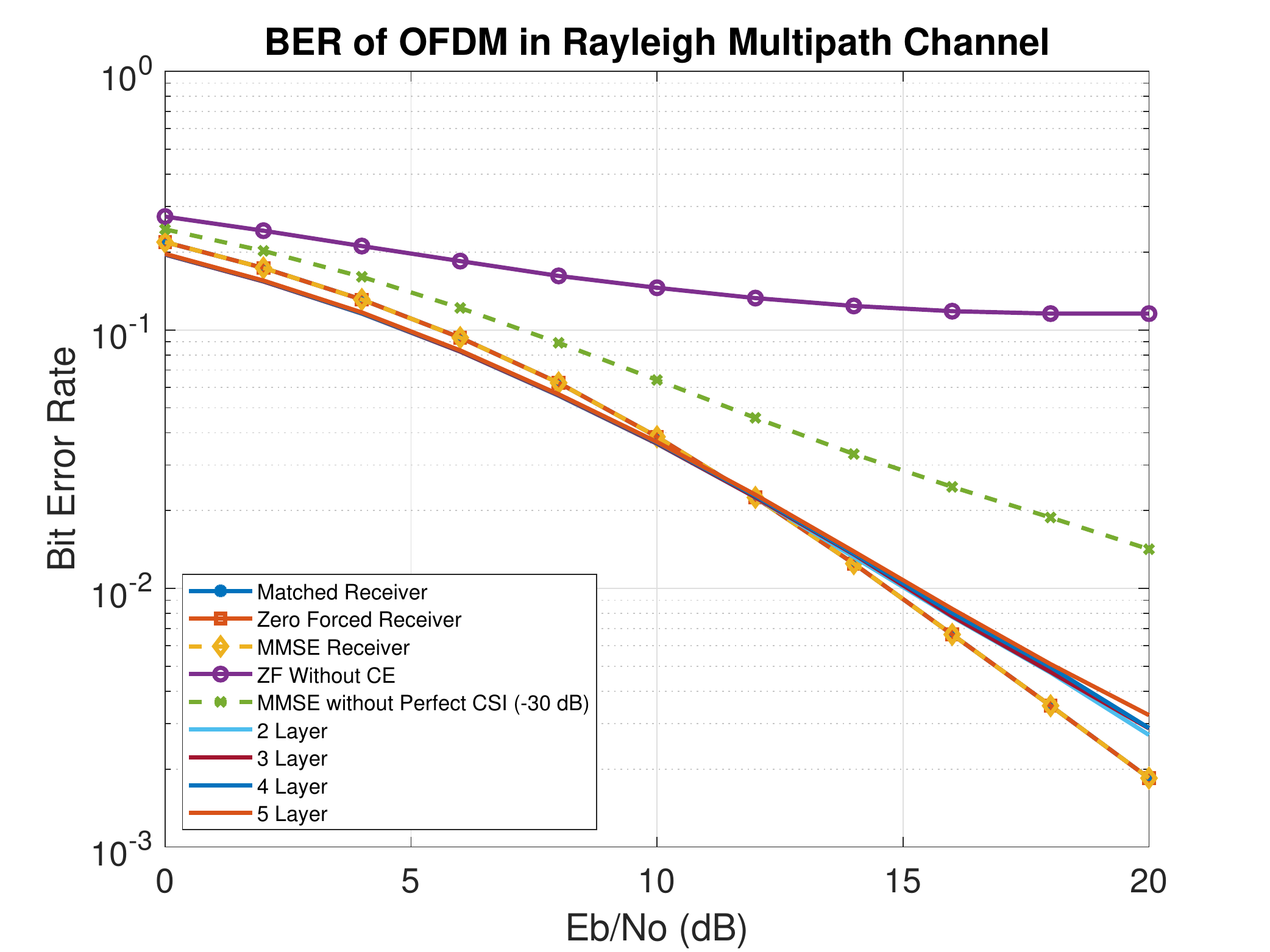}
  \vspace{-0.24in}
  \caption{\small OFDM results for MLP}
  \label{fig:ofdmmlp}
\end{subfigure}
\\
  \begin{subfigure}{.33\linewidth}
  \centering
  \includegraphics[width=\linewidth]{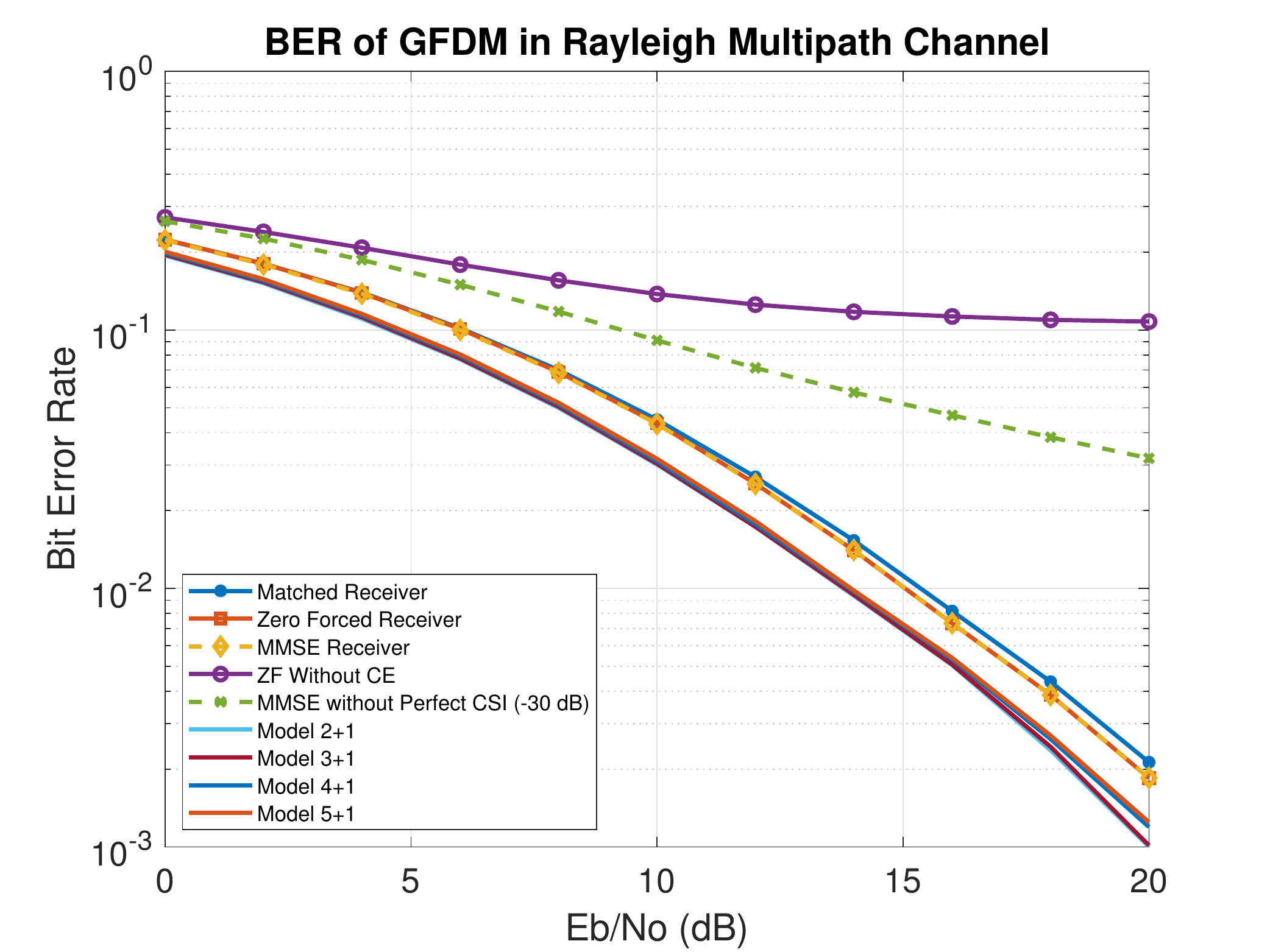}
  \vspace{-0.24in}
  \caption{\small GFDM results for 2D-CNNs}
  \label{fig:gfdm}
\end{subfigure}
~
  \begin{subfigure}{.33\linewidth}
  \centering
  \includegraphics[width=\linewidth]{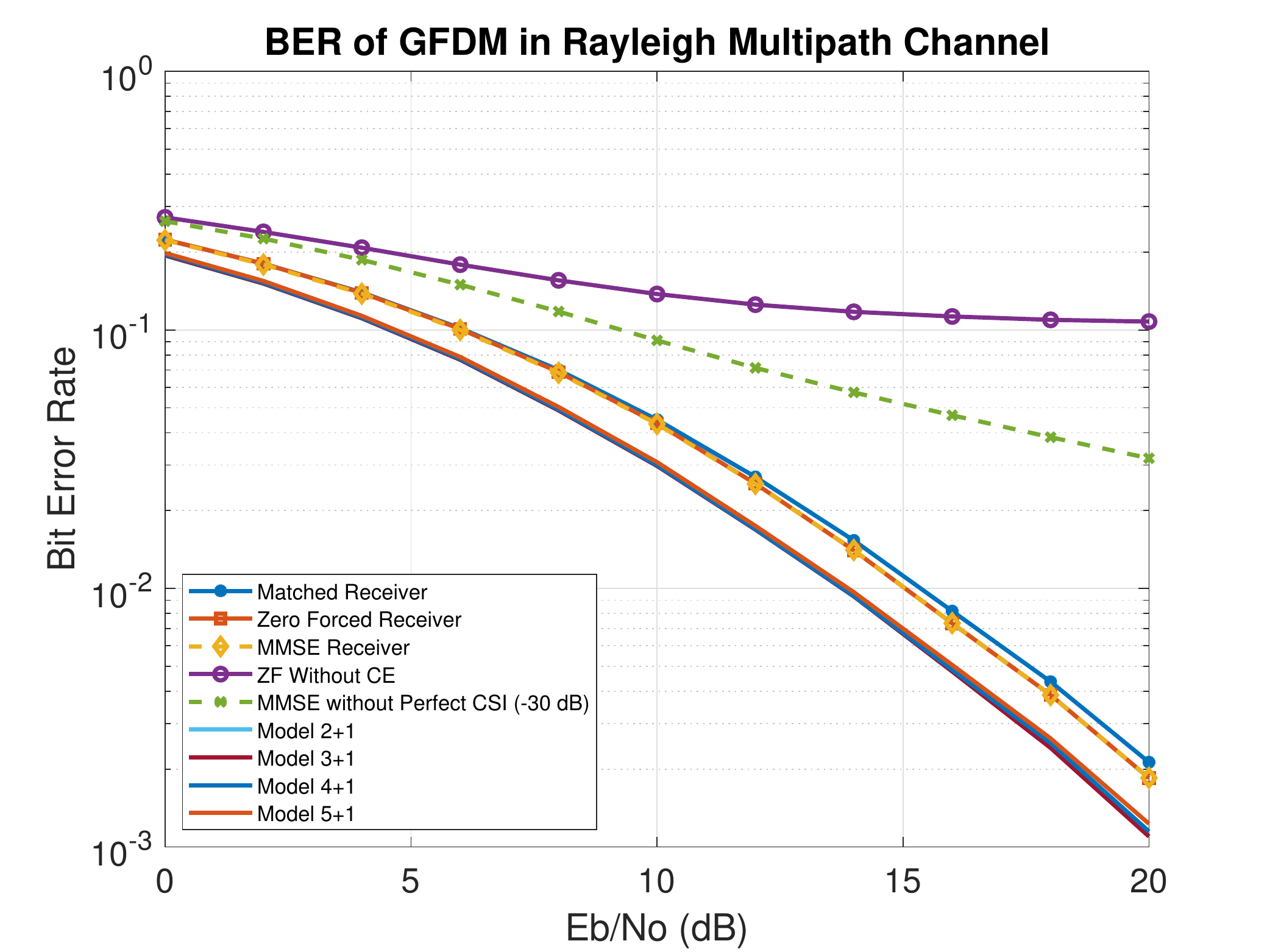}
  \vspace{-0.24in}
  \caption{\small GFDM results for 1D-CNNs}
  \label{fig:GFDM-1Dconv}
\end{subfigure}
~
  \begin{subfigure}{.33\linewidth}
  \centering
  \includegraphics[width=\linewidth]{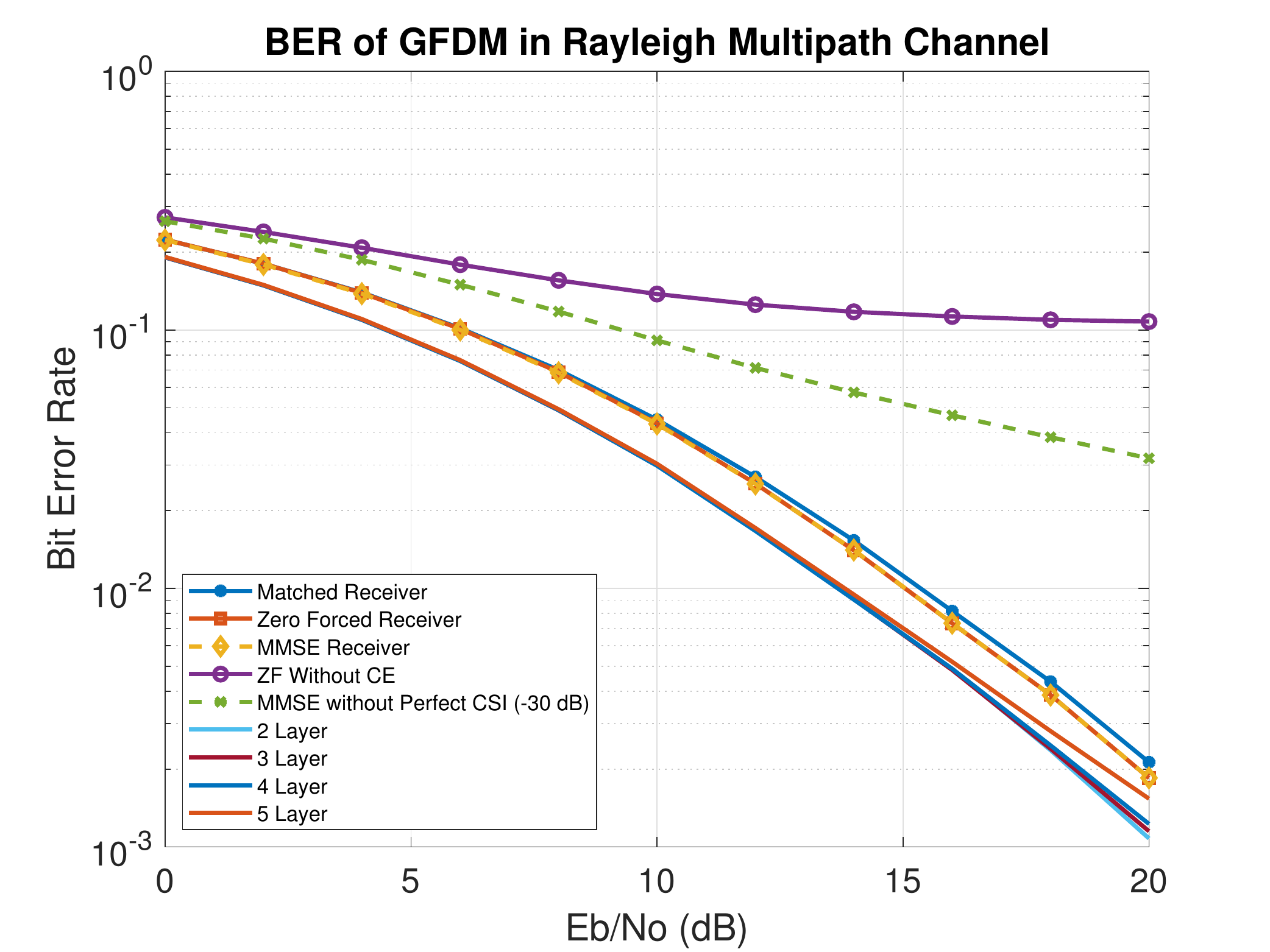}
  \vspace{-0.24in}
  \caption{\small GFDM  results for MLP}
  \label{fig:gfdmmlp}
\end{subfigure}

\vspace{-0.1in}
\caption{\small BER vs. SnR values are shown for both OFDM and GFDM wave-forms using various networks.}
\label{fig:BERFigures}
\end{figure*}

\subsection{Results for OFDM Modulation}

OFDM parameters used in the simulations are given in Table~\ref{gfdm}. First 640000 bits are generated to map 10000 OFDM symbols and splitted over 64 subcarriers ($K$). Cyclic prefix (CP) is inserted at a length of 1 to 4 of the generated time signal. Training data consist of 10000 OFDM symbols and testing data consist of 10000 OFDM symbols. Note that each OFDM symbol carries 64 bits.

We demonstrate how the BER changes with respect to Eb/No (signal to noise ratio) at different  configurations. First, we compare the performance of a 2D-CNN at different layer numbers (where the total number of convolutional layers varies between 2 and 5). The results are summarized in Fig.~\ref{fig:ofdm}. In the figure, the best performance is obtained with the 2+1 2D-CNN. We also demonstrate the performance of another type of convolutional networks: 1D-CNN. Fig.~\ref{fig:OFDM-1Dconv} summarizes the results obtained with different 1D-CNNs. The best result is obtained at 2+1 1D-CNN.

Finally, to compare CNN performance to that of MLP, we study the performance of various MLP networks. The results are shown in Fig.~\ref{fig:ofdmmlp}. The best performance is obtained with 2-layers MLP.

\subsection{Results for GFDM }

GFDM parameters used in the simulations are given in Table \ref{gfdm}. First 960000 bits are generated to map 10000 GFDM symbol and splitted over 32 subcarriers $(K)$ and 3 subsymbols $(M)$. These 32 subcarriers and 3 subsymbols located in time-frequency space forms one GFDM symbol. The chosen pulse shape for the GFDM prototype filter is the root raised cosine (RRC) filter which is widely used in practice with a roll-off factor ($a$) of 0.1. Cyclic prefix (CP) is inserted at a length of 1 to 4 of the generated time signal. Training data consist of 10000 GFDM symbols and testing data consist of 10000 GFDM symbols. Note that each GFDM symbol carries 96 bits.

In the deep receiver, received complex signal with dimensions of 10000x96 is splitted into real and imaginary parts and a represented as 2D data forming 10000x96x2 dimensional 10000 training symbols. The output dimension is 10000x192x1 real vector.

 \addtolength{\topmargin}{+.075in}

We demonstrate how the BER changes with respect to Eb/No (signal to noise ratio) for different  configurations for GFDM. Similar to OFDM experiments, we first compare the performance of a 2D-CNN at different layer numbers (where the total number of convolutional layers varies between 2 and 5).  The  results are summarized in Fig.~\ref{gfdm}. In the figure, the best performance is obtained with the 2+1 2D-CNN. 1D-CNN results are summarized in Fig.~\ref{fig:GFDM-1Dconv}. In the figure, the best result is obtained at 2+1 1D-CNN. As shown in in Fig.~\ref{fig:gfdmmlp}, the best MLP results obtained with 2-layers MLP.

For both OFDM and GFDM simulations, learning rate of Adam optimizer is set to 0.0001. Dropout layer is added to the output of each layer (dropout parameter of 0.1) to avoid overfitting.

When there are some changes in channel taps, whole processing is not fully changed. In classical methods all the processing is re-evaluated. In the deep learning technique, this re-evaluation process can be mitigated with the help of transfer learning. It is seen that it converges even at 1 epoch. Therefore adaptation to the channel is provided faster.

\begin{table}[!t]
\vspace{0.1in}
\caption{Comparison of Trainable Parameters}
\begin{center}
\begin{tabular}{|c|c|c|c|}
\hline
\multicolumn{4}{|c|}{\textbf{Numbers of Trainable Parameters}} \\
\cline{1-4} 
\textbf{Method} & \textbf{\textit{Model Name}}& \textbf{\textit{OFDM}}& \textbf{\textit{GFDM}} \\
\hline
  
 & 2+1 & 132832  & 296736  \\ \cline{2-4} 
Conv1D & 3+1 & 139040  & 302944  \\ \cline{2-4} 
 & 4+1 & 163744  & 327648  \\ \cline{2-4} 
 & 5+1 & 262304  & 426208  \\ \cline{1-4} 
 
 & 2+1 & 134416  & 298320  \\ \cline{2-4} 
Conv2D & 3+1 & 146768  & 310672  \\ \cline{2-4} 
 & 4+1 & 196048  & 359952  \\ \cline{2-4} 
 & 5+1 & 392912  & 556816  \\ \cline{1-4} 

 & 2  & 2130688 & 4752192 \\ \cline{2-4} 
MLP & 3 & 1090368 & 2401152 \\ \cline{2-4} 
 & 4 & 568160  & 1223584 \\ \cline{2-4} 
 & 5 & 306544  & 634288  \\ \cline{2-4}

\hline
\end{tabular}
\label{tab:trainpar}
\end{center}
\end{table}

For the CNN based receiver result, our method approaches the scenario where the channel is fully known. Note that the deep receiver gives results without channel estimation and equalization.

\subsection{Comparison of Parameters}

We compare the total number of parameters for the models used in this paper for 1D-CNNs, 2D-CNNs and MLPs in Table~\ref{tab:trainpar}. Each entry in the table shows the sum of trainable weights and biases used in that particular model. As shown in the table, MLPs typically require more parameters to be trained than CNNs. However, keep in mind that those numbers vary with respect to the total number of filters used in CNNs and the total number of neurons used in each layer in FC layers in MLPs.

\section{Conclusion}
In this paper, we present a deep receiver architecture for multi-carrier wireless systems.
To our best knowledge, this is the first work that introduces data detector without channel equalization using only a deep 2D CNN architecture by eliminating the need for using a coarse detector for multi-carrier systems comparing both OFDM and GFDM.

 \addtolength{\topmargin}{+.075in}
 
In our deep receiver, we analyzed the performance of various neural network architectures including 1D-CNNs, 2D-CNNs and MLPs. Furthermore, we compared those networks' performance to the classical techniques. In our preliminary simulations, we tested networks containing different layers with different hyper-parameters and as shown in our results, shallow networks yielded the best performance (while the difference was not much different, when compared to the deeper architectures). For example, for both OFDM and GFDM experiments, 2+1 networks yielded the best BER among CNN architectures. 

While MLPs can yield slightly better performance for OFDM, they may require significantly more parameters to be kept in the memory. A further study can focus on analyzing the performance of various network architectures on a larger set of multi-carrier forms with larger sets of data.

\bibliographystyle{IEEEtran.bst}
\bibliography{IEEEexample.bib}


\end{document}